\documentclass[runningheads]{llncs}

\usepackage{silence}
\usepackage[font=small, labelfont=bf, justification=raggedright]{caption}

\usepackage[T1]{fontenc}
\usepackage[utf8]{inputenc}
\usepackage{acro}
\usepackage{adjustbox}
\usepackage{booktabs}
\usepackage{comment}
\usepackage{todonotes}
\usepackage{colortbl}
\usepackage{hyperref}

\newcolumntype{a}{>{\columncolor{gray!10!white}}c}
\newcolumntype{x}{>{\columncolor{green!10!white}}c}
\newcolumntype{y}{>{\columncolor{blue!10!white}}c}
\newcolumntype{z}{>{\columncolor{yellow!10!white}}c}
\newcolumntype{v}{>{\columncolor{red!10!white}}c}
\definecolor{OliveGreen}{rgb}{0,0.6,0}
\definecolor{ForestGreen}{RGB}{34,139,34}
\definecolor{myblue}{RGB}{37,165,203}
\definecolor{FAUblue}{rgb}{0.000, 0.2196, 0.3961}
\definecolor{myred}{RGB}{175,32,67}

\colorlet{backgroundcol}{cyan!10!white}

\usepackage{graphicx}
\usepackage{enumitem}
\usepackage{color}
\usepackage{xcolor}
\usepackage[table]{xcolor}
\usepackage{listings}
\definecolor{codegreen}{rgb}{0,0.6,0}
\definecolor{codegray}{rgb}{0.5,0.5,0.5}
\definecolor{codepurple}{rgb}{0.58,0.5,0.82}
\definecolor{backcolour}{rgb}{0.95,0.95,0.92}

\lstdefinestyle{mystyle}{
    backgroundcolor=\color{white},   
    commentstyle=\color{codegreen},
    keywordstyle=\color{magenta}\bfseries,
    moredelim=[is][\color{magenta}\bfseries]{@}{@},
    numberstyle=\tiny\color{codegray},
    stringstyle=\color{codepurple},
    basicstyle=\C\footnotesize,
    frame           = tb,         % <-- draws TOP and BOTTOM rules
    framerule       = 0.6pt,      % <-- thickness of the rules
    rulecolor       = \color{black},
    framesep        = 0.4em,      % space between rule and code
    xleftmargin     = 2em,
    framexleftmargin= 2em,
    breakatwhitespace=false,         
    breaklines=true,                 
    captionpos=b,                    
    keepspaces=true,                 
    numbers=left,                    
    numbersep=5pt,                  
    showspaces=false,                
    showstringspaces=false,
    showtabs=false,                  
    tabsize=2
}

\usepackage{amsmath,amssymb,amsfonts}

\usepackage{lipsum} 
\usepackage{graphicx}
\usepackage{soul}

\usepackage{adjustbox}

\usepackage{tabularx}
\usepackage{pifont}

\begin{document}
%%----------------

% \title{Heterogeneous Hybrid MPI+OpenMP PIC Monte Carlo Simulations with Persistent GPU Memory, Runtime Interoperability and Asynchronous Multi-GPU Execution}

\title{Multi-GPU Hybrid Particle-in-Cell Monte Carlo Simulations for Exascale Computing Systems}

%
%\titlerunning{Abbreviated paper title}
\titlerunning{Multi-GPU Hybrid PIC MC Simulations for Exascale Computing Systems}
% If the paper title is too long for the running head, you can set
% an abbreviated paper title here
%
% \author{Jeremy J. Williams\thanks{Corresponding location: Lindstedtsvägen 5, SE-100 44, Stockholm, Sweden \\ E-mail address: jjwil@kth.se (Jeremy J. Williams)}\inst{1}\and}

\author{Jeremy J. Williams\inst{1}\and
Jordy Trilaksono \inst{2} \and
Stefan Costea\inst{3} \and
Yi Ju \inst{6} \and
Luca Pennati\inst{1} \and
Jonah Ekelund\inst{1} \and
David Tskhakaya \inst{4} \and
Leon Kos \inst{3} \and
Ales Podolnik \inst{4} \and
Jakub Hromadka\inst{4} \and
Allen D. Malony\inst{5} \and
Sameer Shende\inst{5} \and
Tilman Dannert \inst{6} \and
Frank Jenko \inst{2} \and
Erwin Laure \inst{6} \and
Stefano Markidis\inst{1} }

\authorrunning{Jeremy J. Williams et al.}
% First names are abbreviated in the running head.
% If there are more than two authors, 'et al.' is used.
%
\institute{KTH Royal Institute of Technology, Stockholm, Sweden \and
Max Planck Institute for Plasma Physics, Garching, Germany \and
Faculty of Mechanical Engineering, University of Ljubljana, Ljubljana, Slovenia \and
Institute of Plasma Physics of the CAS, Prague, Czech Republic  \and 
University of Oregon, Eugene, Oregon, USA \and
Max Planck Computing and Data Facility, Garching, Germany}
\maketitle              % typeset the header of the contribution
\begin{abstract}
Particle-in-Cell (PIC) Monte Carlo (MC) simulations are central to plasma physics but face increasing challenges on heterogeneous HPC systems due to excessive data movement, synchronization overheads, and inefficient utilization of multiple accelerators. In this work, we present a portable, multi-GPU hybrid MPI+OpenMP implementation of BIT1 that enables scalable execution on both Nvidia and AMD accelerators through OpenMP target tasks with explicit dependencies to overlap computation and communication across devices. Portability is achieved through persistent device-resident memory, an optimized contiguous one-dimensional data layout, and a transition from unified to pinned host memory to improve large data-transfer efficiency, together with GPU Direct Memory Access (DMA) and runtime interoperability for direct device-pointer access. Standardized and scalable I/O is provided using openPMD and ADIOS2, supporting high-performance file I/O, in-memory data streaming, and in-situ analysis and visualization. Performance results on pre-exascale and exascale systems, including Frontier (OLCF-5) for up to 16,000 GPUs, demonstrate significant improvements in run time, scalability, and resource utilization for large-scale PIC MC simulations.


\end{abstract}

% ========================
% # Introduction      # 
% ========================
\section{Introduction}
High-performance computing (HPC) is rapidly moving toward heterogeneous architectures~\cite{milojicic2021future}, where efficient orchestration of computation, memory management, and overlap of communication and computation are essential for large-scale particle-based plasma simulations on pre-exascale and exascale systems~\cite{vasileska2020modernization}.

BIT1 is a massively parallel Particle-in-Cell (PIC) Monte Carlo (MC) code for plasma simulations and plasma–material interactions. While earlier MPI-only versions scaled well on CPU clusters~\cite{tskhakaya2010pic}, moving to GPU supercomputers exposed limitations in on-node communication, memory usage, and efficient multi-GPU utilization. To address this, BIT1 was extended with hybrid MPI+OpenMP and MPI+OpenACC parallelization for shared-memory execution and initial GPU acceleration, which also identified key bottlenecks such as the particle mover and arranger~\cite{williams2023leveraging,williams2024optimizing,williams2025accelerating}, and was complemented by scalable data management and I/O based on the openPMD standard with parallel~\cite{williams2024enabling}, streaming~\cite{williams2026integrating}, and in-situ workflows~\cite{williams2024understanding} to reduce storage overhead and post-processing costs.

In this work, we present a portable multi-GPU hybrid MPI\allowbreak+OpenMP BIT1 implementation with persistent device-resident memory, optimized data transfers, GPU runtime interoperability for seamless integration between CUDA, HIP, and OpenMP runtimes, and asynchronous multi-GPU execution. By keeping data resident on the GPU, minimizing redundant transfers, and coordinating kernels asynchronously across devices, BIT1 efficiently exploits modern GPU accelerated systems while maintaining large-scale MPI scalability and portability across Nvidia and AMD GPU architectures.

The main contributions of this work are:

\begin{itemize}
    \item We design hybrid MPI+OpenMP versions of BIT1 to improve performance on a single node and in both strong and weak scaling tests, employing a task-based approach to mitigate load imbalance and optimize resource utilization.
    \item We extend OpenMP GPU offloading for Nvidia and AMD, enabling portability using OpenMP target tasks with \texttt{nowait} and \texttt{depend} clauses, overlapping computation and communication, flattening 3D arrays into contiguous 1D layouts, and maintaining device-resident memory across time steps using OpenMP target \texttt{enter\allowbreak/exit data} regions, and switching from unified to pinned host memory for large arrays to reduce data movement overheads.
    \item We implement vendor-specific GPU optimizations to improve data-transfer efficiency, using compile-time pinned memory on Nvidia and OpenMP pinned allocators on AMD, enabling higher throughput and lower-latency host-to-device (HtoD) transfers for large arrays.
    \item We integrate openPMD and ADIOS2 (BP4 and SST backends) to remove I/O bottlenecks, enabling high-throughput I/O and in-memory streaming, with in-situ analysis and visualization for real-time insight and reduced post-processing on Nvidia and AMD GPUs architectures at scale.
\end{itemize}

\section{Background}
The PIC method is one of the most widely used computational approaches for plasma simulations and is applied across a broad range of plasma environments, including space and astrophysical plasmas, laboratory experiments, industrial processes, and fusion devices~\cite{tskhakaya2007particle,williams2023leveraging}. 

% \begin{wrapfigure}{r}{0.55\textwidth}
%  \vspace{-0.6cm} % Adjust vertical figure placement
% \centering
% \includegraphics[width=1\linewidth]{images/PIC_Method_Diagram_With_Focus_Particle_Mover_Gray.png}
% \caption{A diagram representing the PIC Method on HPC architectures. After initialization, the PIC method repeats at each time step. In gray, we highlight the particle mover step that we parallelize in the portable multi-GPU hybrid BIT1.}
% \label{fig:pic_method_cycle}
% \vspace{-0.6cm} % Adjust vertical figure spacing
% \end{wrapfigure} 

\begin{figure*}[!ht]
    \vspace{-0.4cm} % Adjust vertical figure placement
    \begin{center}
        \includegraphics[width=0.55\textwidth]{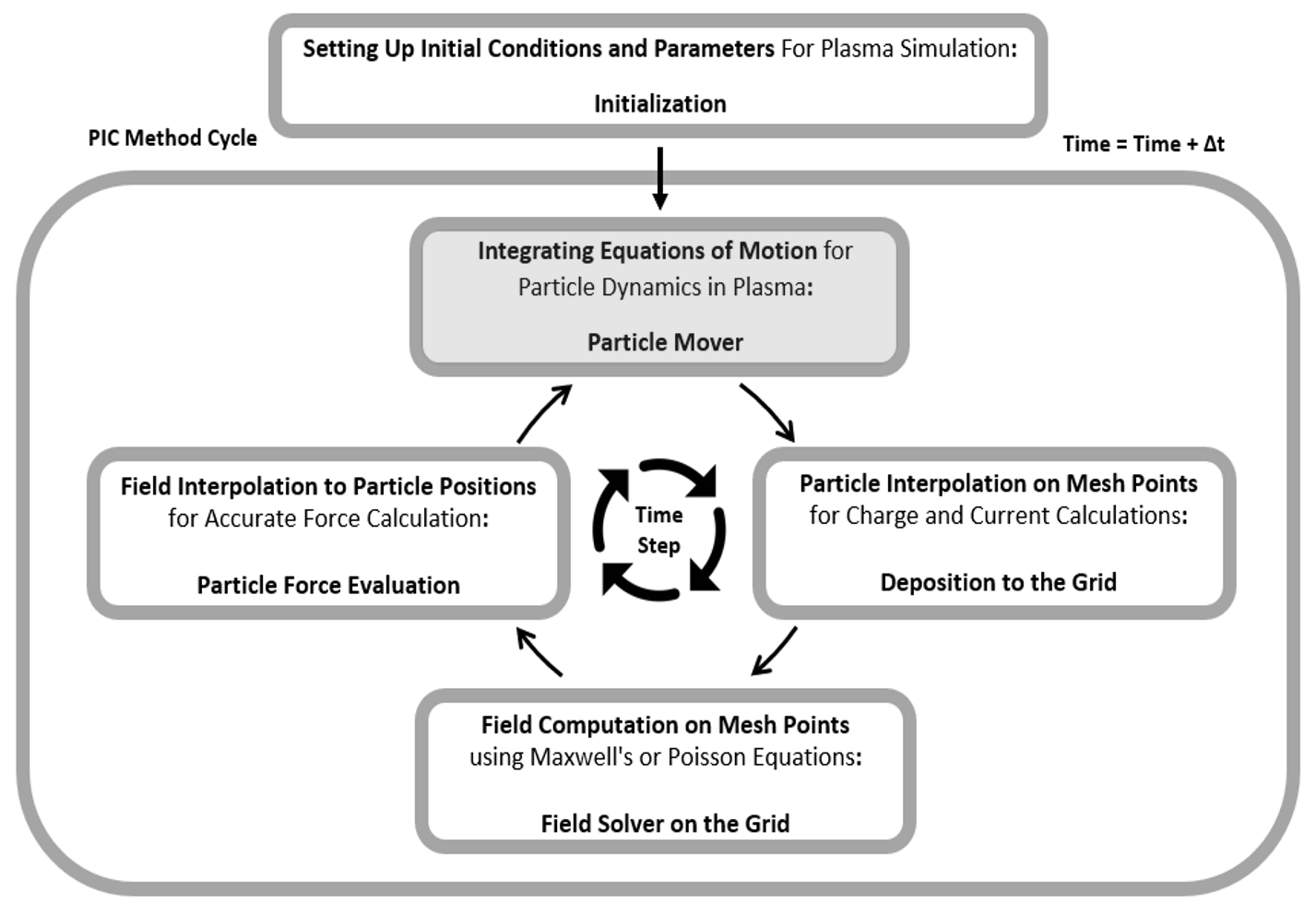}
        \caption{A diagram representing the PIC Method on HPC architectures. After initialization, the PIC method repeats at each time step. In gray, we highlight the particle mover step that we parallelize in the portable multi-GPU hybrid BIT1.} 
        \label{fig:pic_method_cycle}
     \end{center}
     \vspace{-0.6cm} % Adjust vertical figure placement
\end{figure*}

As seen in Fig.~\ref{fig:pic_method_cycle}, the classical PIC method begins with an initialization stage, in which the simulation setup and initial conditions are defined. The main simulation loop then executes four phases: (i) the particle mover, which updates particle positions and velocities; (ii) deposition to the grid, which maps particle charges and currents to the mesh; (iii) the field solver, which computes electromagnetic or electrostatic fields by solving Maxwell’s or Poisson’s equations; and finally (iv) particle force evaluation, which interpolates the grid fields back to the particle positions. This cycle is repeated for each time step, with each step representing a small fraction of the characteristic plasma timescale in order to accurately capture plasma dynamics~\cite{williams2023leveraging,williams2025accelerating}.

% In this work, we use the Berkeley Innsbruck Tbilisi 1D3V (BIT1) code, an application tool designed for large-scale plasma simulations on HPC systems and tailored to modeling plasma edge and tokamak divertor physics~\cite{tskhakaya2007optimization,tskhakaya2010pic}. BIT1 is a specialized one-dimensional, three-velocity (1D3V) electrostatic PIC MC code, derived from the \texttt{XPDP1} code developed by Birdsall’s group at the University of California, Berkeley, in the 1990s~\cite{verboncoeur1993simultaneous}. Written entirely in C and comprising approximately 31,600 lines of code, BIT1 relies on native implementations of the Poisson solver, particle mover, and smoothing operators, without dependencies on external numerical libraries. BIT1 employs the "natural sorting" method~\cite{tskhakaya2007optimization}, which accelerates collision operators by up to a factor of five, but increases memory requirements and introduces challenges for GPU offloading and load balance in regions of high particle density~\cite{williams2023leveraging,williams2025accelerating}. To address these limitations and ensure portability across modern supercomputing platforms, BIT1 is evolving toward a hybrid parallel programming model that combines MPI-based domain decomposition with OpenMP shared-memory parallelism, reducing communication overhead and improving scalability on multi-core and heterogeneous architectures~\cite{williams2024optimizing,williams2025accelerating}.

In this work, we use the Berkeley Innsbruck Tbilisi 1D3V (BIT1) code for large-scale plasma simulations on HPC systems, targeting plasma edge and tokamak divertor physics~\cite{tskhakaya2007optimization,tskhakaya2010pic}. BIT1 is a 1D3V electrostatic PIC MC code derived from Birdsall’s \texttt{XPDP1} code (UC Berkeley, 1990s)~\cite{verboncoeur1993simultaneous}. It is written in C (~31,600 lines of code) and uses native Poisson solver, particle mover, and smoothing operators without external numerical libraries.

BIT1 employs the "natural sorting" method~\cite{tskhakaya2007optimization}, which accelerates collision operators by up to a factor of five, but increases memory use and complicating GPU offloading and load balancing in high-density regions~\cite{williams2023leveraging,williams2025accelerating}. To address these limitations and improve portability, BIT1 is evolving toward a hybrid MPI + OpenMP model that reduces communication overhead and improves scalability on multi-core and heterogeneous architectures~\cite{williams2024optimizing,williams2025accelerating}.
 
\subsection{OpenMP Memory Allocators, Unified \& Pinned Host Memory}
OpenMP (Open Multi-Processing) provides a standardized mechanism for controlling memory allocation through \emph{memory allocators} (Table~\ref{tab:predefined_allocators}), which return logically contiguous memory from a specified memory space and guarantee that allocations do not overlap. The behavior of an allocator is configured using \emph{allocator traits} (Table~\ref{tab:allocator_traits}) that define synchronization expectations, alignment, accessibility, fallback behavior, pinning and memory placement, providing a portable abstraction for heterogeneous memory hierarchies without relying on vendor-specific API capabilities. 

\begin{table}[!ht]
% \vspace{-0.4cm} % Adjust vertical figure placement
\centering
\scriptsize
\begin{tabularx}{\linewidth}{p{4cm}p{4cm}X}
\hline
\textbf{OpenMP Allocator Name} & \textbf{Associated Memory Space} & \textbf{Primary Use} \\
\hline
\rowcolor{gray!20}
\texttt{omp\_default\_mem\_alloc} &
\texttt{omp\_default\_mem\_space} &
Default system storage \\
\hline
\end{tabularx}
\vspace{2mm}  % adjust vertical spacing below
\caption{Predefined OpenMP allocator and memory space.~\cite{neth2021beyond,sewall2016modern}}
\label{tab:predefined_allocators}
\vspace{-0.4cm} % Adjust vertical figure placement
\end{table}

\begin{table}[!ht]
% \vspace{-0.6cm} % Adjust vertical figure placement
\centering
\scriptsize
\begin{tabularx}{\linewidth}{p{4cm}p{5cm}X}
\hline
\textbf{OpenMP Allocator Trait} & \textbf{Allowed Values} & \textbf{Default Setting} \\
\hline
\rowcolor{gray!20}
\texttt{pinned} &
\texttt{true}, \texttt{false} &
\texttt{false} \\
\hline
\end{tabularx}
\vspace{2mm}
\caption{OpenMP allocator trait, accepted values, and default setting.~\cite{neth2021beyond,sewall2016modern}}
\label{tab:allocator_traits}
\vspace{-0.6cm} % Adjust vertical figure placement
\end{table}

Efficient data movement is critical for particle-based PIC codes on heterogeneous systems. Unified (managed) memory (UM) offers a single virtual address space with on-demand page migration between CPU and GPU, simplifying programming but potentially incurring overheads due to page faults and runtime migrations for irregular or latency-sensitive access patterns~\cite{mishra2017benchmarking}. In contrast, pinned (page-locked) host memory (PinM) enables higher-bandwidth and lower-latency transfers, supports asynchronous copies and GPU DMA, but must be managed explicitly and can increase pressure on host memory resources. For performance critical PIC MC kernels, such as particle movers and collision operators, PinM is therefore commonly preferred when explicit data movement and overlap of communication and computation are required~\cite{choi2021comparing,noaje2010multigpu}.

\subsection{openPMD Standard, openPMD-api Integration \& ADIOS2}
The openPMD (open Particle-Mesh Data) standard defines a portable, extensible data model for particle and mesh data, supporting multiple storage backends such as HDF5, ADIOS1, ADIOS2, and JSON in serial and MPI workflows~\cite{openPMDstandard}. The openPMD-api library provides a unified interface, storing physical quantities as multi-dimensional records for mesh fields or particle data, with time-dependent data organized as iterations~\cite{openPMDapi}.

ADIOS2 (Adaptable Input/Output System version 2) is an open-source parallel I/O framework for scalable HPC data movement, offering a unified API for multi-dimensional variables, attributes, and time steps, with modular engines including BP4 (high-throughput file I/O) and SST (low-latency streaming for in-situ MPI-parallel workflows)~\cite{williams2026integrating}. In hybrid BIT1, we use BP4 for high-throughput file I/O~\cite{williams2024enabling,williams2024understanding} and SST for in-memory data streaming~\cite{williams2026integrating} independently, enabling in-situ analysis and visualization without interrupting asynchronous multi-GPU execution on pre-exascale and exascale systems.

% ==================
% # Related Work #
% ==================
\section{Related Work}
The rapid transition of scientific simulations toward exascale has driven a shift from traditional CPU implementations to heterogeneous, accelerator focused, and hybrid programming models capable of efficiently leveraging multiple GPU devices. Early work in hybrid parallelization for particle-based simulations explored OpenMP and MPI for scalable performance on HPC clusters~\cite{chaudhury2018hybrid}. GPU programming models have evolved significantly, with OpenMP offloading emerging as a promising framework for portability across Nvidia, AMD, and Intel GPUs~\cite{mehta2020evaluating}. Studies have evaluated OpenMP offloading for computational kernels and data-transfer strategies, demonstrating performance comparable to vendor optimized numerical libraries on H100 and MI250X GPUs~\cite{krishnaamy2025openmp}. Other investigations have focused on portable GPU runtimes, with Tian et al.~\cite{tian2021experience} showing that OpenMP 5.1, with minor compiler extensions, can replace vendor-specific runtimes such as CUDA or HIP without performance loss, enabling efficient compilation across LLVM/Clang toolchains for multiple GPU architectures. Similarly, portable implementations of MC particle transport codes, such as OpenMC, have been successfully ported to heterogeneous GPUs using OpenMP target offloading, with large-scale benchmarks confirming efficiency across AMD, Nvidia, and Intel accelerators~\cite{tramm2022toward}. 

% ====================
% # METHODOOGY #
% ====================

\section{Methodology \& Experimental Setup}
In this work, we focus on a portable, multi-GPU hybrid MPI\allowbreak+OpenMP implementation of BIT1 for PIC MC simulations on exascale systems, leveraging persistent GPU memory, contiguous 1D data layouts, runtime interoperability, asynchronous multi-GPU execution, and OpenMP target offloading with vendor-specific optimizations, while integrating high-performance file and streaming I/O, and in-situ workflows using openPMD and ADIOS2.

\subsection{Portable Multi-GPU Hybrid BIT1}
OpenMP is a widely adopted programming model for shared-memory parallelism in HPC, supported by major compilers including GCC, LLVM, and Intel, providing portability across platforms and simplifying the development of parallel applications through directives and runtime routines.

\smallskip
\noindent\textbf{OpenMP Target Parallelization, Data Structure \& Memory Layout.}
\noindent Previously, Williams et al.~\cite{bit1openmptasksmover,williams2024optimizing,williams2025accelerating} developed an OpenMP target tasks implementation for the particle mover in hybrid BIT1. In this function, particle positions and velocities are stored in \texttt{x[species]\allowbreak[cell]\allowbreak[particle]} and \texttt{vx[species]\allowbreak[cell]\allowbreak[particle]}, where \texttt{nsp}, \texttt{nc} and \texttt{np[species]\allowbreak[cell]} denote the number of species, cells and particles per cell, respectively~\cite{tskhakaya2007optimization}. Such 3D data layouts cause non-contiguous GPU accesses~\cite{wu2013complexity}, reducing memory efficiency and bandwidth. To improve GPU runtime performance, hybrid BIT1 adopts a revised data layout after identifying limitations in large 3D arrays, such as \texttt{x[species]\allowbreak[cell]\allowbreak[particle]}. These large arrays were restructured into contiguous 1D arrays (\texttt{x\_GPU\allowbreak[species\allowbreak*cell\allowbreak*particle]} or \texttt{x\_GPU\allowbreak[:LenA]}) to enable sequential memory access, simpler indexing and higher sustained throughput in GPU kernels. HtoD data movement is optimized using \texttt{PinM} to ensure fixed physical data residency, supporting efficient asynchronous transfers and GPU DMA. On Nvidia GPUs, \texttt{PinM} is selected at compile-time, while on AMD GPUs, where compile-time UM/PinM is unavailable, OpenMP memory allocators explicitly manage memory placement and selective pinning of performance critical arrays, minimizing overhead while maintaining portability across heterogeneous GPU architectures.

{\centering
\begin{minipage}{0.73\textwidth}
% \vspace{-0.1cm} % Adjust vertical figure placement
\begin{lstlisting}[
    language=C,
    style=mystyle,
    basicstyle=\fontsize{4}{4}\selectfont\ttfamily,
    caption={Simplified C code snippet illustrating the Persistent Asynchronous GPU OpenMP Offloading},
    captionpos=b,
    label=1st:openmp_target_persistent_snippet,
    tabsize=2,
    lineskip=0pt,
    xleftmargin=2em, 
    framexleftmargin=2em
]
// Enter unstructured OpenMP target data region
@#pragma@ omp target enter data map(to: chsp[:lenA], sn2d[:lenA], dinj[:lenA], 
poff[:lenA], nstep[:lenA], maxL[:lenA], np_GPU[:lenA], x_GPU[:lenA], 
y_GPU[:lenA], vx_GPU[:lenA], vy_GPU[:lenA])
// Persistent device-resident (GPU) memory allocation
{
    while (tstep < last_step) {
        ...
        ...
        (*move_p)(); // Particle mover -> move0, moveb     
        ...
        ...
        xarrj(); // Particle arranger -> arrj, narrj 
        ...
        ...
        // I/O -> openPMD, ADIOS2, dumpstep, datfiles 
        ...
        ...
    } // End of while(tstep < last_step)
 ...
}
// Exit unstructured OpenMP target data region
@#pragma@ omp target exit data map(delete: x_GPU[:lenA], y_GPU[:lenA])
// Release persistent device-resident (GPU) memory
\end{lstlisting}
\vspace{-0.1cm} % Adjust vertical figure placement
\end{minipage}
\par}

\smallskip
% \subsubsection{Persistent Device-Resident Memory Allocation.}
\noindent\textbf{Persistent Device-Resident Memory Allocation.}
\noindent Building on the asynchronous GPU acceleration of the particle mover~\cite{williams2025accelerating}, hybrid BIT1 reduces GPU data movement overhead by using persistent device-resident memory allocation. In Listing~\ref{1st:openmp_target_persistent_snippet}, rather than repeatedly transferring and allocating large arrays (\texttt{x\_GPU\allowbreak[:LenA]}, \texttt{vx\_GPU\allowbreak[:LenA]}, \texttt{y\_GPU\allowbreak[:LenA]}, \texttt{vy\_GPU\allowbreak[:LenA]}) each time step, a single \texttt{\#pragma \allowbreak omp \allowbreak target \allowbreak enter \allowbreak data} region persistently allocates all relevant arrays and control data (\texttt{chsp}, \texttt{sn2d}, \texttt{dinj}, \texttt{poff}, \texttt{nstep}, \texttt{maxL}, \texttt{np\_GPU}) on the GPU. The main simulation loop then iterates over time steps, executing the particle mover (\texttt{move0, moveb}), particle arranger (\texttt{arrj, narrj}), and performing in-situ I/O without repeated data transfers. After the simulation, \texttt{\#pragma} \allowbreak \texttt{omp} \allowbreak \texttt{target} \allowbreak \texttt{exit} \allowbreak \texttt{data} \allowbreak \texttt{map(delete: ...)} releases the device-resident memory.

{\centering
\begin{minipage}{0.68\textwidth}
% \vspace{-0.3cm} % Adjust vertical figure placement
\begin{lstlisting}[
    language=C,
    style=mystyle,
    basicstyle=\fontsize{4}{4}\selectfont\ttfamily,
    caption={Simplified C code snippet illustrating OpenMP memory allocators with the PinM trait enabled.},
    captionpos=b,
    label=2nd:openmp_memory_allocators_snippet,
    tabsize=2,
    lineskip=0pt,
    xleftmargin=2em, 
    framexleftmargin=2em
]
/* Initialize OMP Pinned Allocator */
omp_allocator_handle_t pinned_alloc = omp_null_allocator;
...
// Create Pinned OpenMP Memory Allocator
if (pinned_alloc == omp_null_allocator) {
	omp_alloctrait_t traits[1];
	...
	pinned_alloc = omp_init_allocator(omp_default_mem_space, 1, traits);
	...
}
// Pinned Allocation
x_GPU = (float *) omp_alloc(lenA * sizeof(float), pinned_alloc);
...
// Persistent Device-Resident (GPU) Memory Allocation
...
// GPU OpenMP Offloading, Sorting, Computation, I/O 
...
// Release Persistent Device-Resident (GPU) Memory
...
// Final I/O -> openPMD, ADIOS2, dumpstep, datfiles
...
// Destroy and Cleanup OpenMP Pinned Memory
if (pinned_alloc != omp_null_allocator) {
    omp_free(x_GPU,  pinned_alloc);
	...
	omp_destroy_allocator(pinned_alloc);
	pinned_alloc = omp_null_allocator;
}
...
\end{lstlisting}
\vspace{-0.1cm} % Adjust vertical figure placement
\end{minipage}
\par}

\smallskip
% \subsubsection{Predefined OpenMP Memory Allocators.}
\noindent\textbf{Predefined OpenMP Memory Allocators.} 
\noindent Efficient data movement is critical in GPU accelerated hybrid BIT1, especially for large particle arrays. Listing~\ref{2nd:openmp_memory_allocators_snippet} shows how hybrid BIT1 reduces repeated allocations and transfers (particularly on AMD GPUs) by using OpenMP \texttt{PinM} memory allocators, which provide fixed residency, high-throughput asynchronous transfers, and GPU DMA while preserving portability. The allocator is initialized once at startup, allocating large arrays (\texttt{x\_GPU}, \texttt{y\_GPU}, \texttt{vx\_GPU}, \texttt{vy\_GPU}) persistently on the GPU for all time steps, enabling asynchronous offloading with overlapping computation and I/O. At simulation end, arrays are released, \texttt{PinM} is freed via \texttt{omp\_free}, and the allocator destroyed with \texttt{omp\_destroy\_allocator}, ensuring clean memory management.

{\centering
\begin{minipage}{0.85\textwidth}
% \vspace{-0.3cm} % Adjust vertical figure placement
\begin{lstlisting}[
    language=C,
    style=mystyle,
    basicstyle=\fontsize{4}{4}\selectfont\ttfamily,
    caption={Simplified C code snippet illustrating OpenMP Target Tasks with "nowait" and "depend" clauses, and enabling GPU runtime interoperability using "use\_device\_ptr" and "is\_device\_ptr" clauses for direct device-pointer access.},
    captionpos=b,
    label=3rd:openmp_gpu_runtime_snippet, 
    tabsize=2,
    xleftmargin=2em, 
    framexleftmargin=2em 
]
// Host pointer already mapped into its device pointer
@#pragma@ omp target data use_device_ptr(np_GPU, x_GPU, y_GPU, vx_GPU, vy_GPU)
{
  for (isp=0; isp<nsp; isp++) { 
      ...     
      @#pragma@ omp target teams distribute parallel for
      // Do not map; pointer already device-resident
      @is_device_ptr@(poff, nstep, maxL, np_GPU, x_GPU, vx_GPU)
      depend(inout: x_GPU[:lenA])
      depend(in: poff[:lenA], nstep[:lenA], maxL[:lenA], vx_GPU[:lenA], np_GPU[:lenA]) 
      firstprivate(ci0) private(i, j, idx, idx0, cij) 
      thread_limit(512) num_teams(391) nowait
          for(j = 0; j < nc; j++) {
            idx0 = poff[isp] + j*maxL[isp];
            cij = ci0 + j;
            @#pragma@ omp simd
              for (i=0; i< np_GPU[cij]; i++) {
                idx = idx0 + i;
                x_GPU[idx] += nstep[isp]*vx_GPU[idx];
              }
           } 
      ...
  } else {
        ...
    ...
} // end of #pragma omp target data region

// Wait for all asynchronous OpenMP tasks (GPU kernels)
#pragma omp taskwait
\end{lstlisting}
\vspace{-0.1cm} % Adjust vertical figure placement
\end{minipage}
\par}

\smallskip
% \subsubsection{Enable GPU Runtime Interoperability.}
\noindent\textbf{Enable GPU Runtime Interoperability.} 
\noindent Following the transition from UM to PinM to optimize data transfers, hybrid BIT1 further enables GPU runtime interoperability via direct device-pointer access using OpenMP \texttt{use\_device\_ptr} and \texttt{is\_device\_ptr} clauses. In Listing~\ref{3rd:openmp_gpu_runtime_snippet},  the \texttt{\#pragma} \allowbreak \texttt{omp} \allowbreak \texttt{target} \allowbreak \texttt{data} \allowbreak \texttt{use\_device\_ptr} \allowbreak \texttt{(x\_GPU,} \allowbreak \texttt{y\_GPU,} \allowbreak \texttt{vx\_GPU,} \allowbreak \texttt{vy\_GPU)} region informs the runtime that the host pointers are already mapped to device memory, allowing GPU kernels to access device-resident data directly without additional mapping or allocation. Within this data region, GPU kernels are launched with \texttt{\#pragma} \allowbreak \texttt{omp} \allowbreak \texttt{target} \allowbreak \texttt{teams} \allowbreak \texttt{distribute} \allowbreak \texttt{parallel} \allowbreak \texttt{for} \allowbreak \texttt{is\_device\_ptr \allowbreak (...)} so that arrays such as \texttt{poff}, \texttt{nstep}, \texttt{maxL}, \texttt{np\_GPU}, \texttt{x\_GPU}, \texttt{y\_GPU}, \texttt{vx\_GPU}, and \texttt{vy\_GPU} are used as device pointers. The GPU kernels are issued asynchronously using \texttt{nowait} and are ordered through explicit \texttt{depend \allowbreak (in/out/inout)} clauses, with a final \texttt{\#pragma \allowbreak omp \allowbreak taskwait} ensuring that all OpenMP tasks (GPU kernels) are completed. 

% ==========================
% # Experimental Setup #
% ==========================

% \space
\subsection{Use Case \& Experimental Environment}
We focus on evaluating the performance of hybrid BIT1 on two test cases that differ in problem size and BIT1 functionality. We consider two cases: \emph{i)} a relatively straightforward run simulating neutral particle ionization due to interactions with electrons, and \emph{ii)} the formation of a high-density sheath in front of so-called \emph{divertor} plates in future magnetic confinement fusion devices, such as the ITER and DEMO fusion devices. 

More precisely, the two cases are as follows:

\begin{itemize}[leftmargin=*]
\item \textbf{Neutral Particle Ionization Simulation.} An unbounded, unmagnetized plasma of electrons, $D^+$ ions, and $D$ neutrals is considered. Neutral concentration decreases via $\partial n / \partial t = n n_e R$, with $n$, $n_e$, and $R$ the neutral density, plasma density, and ionization rate coefficient. We use a 1D geometry with 100K cells, three species, and initial 100 particles per cell per species (total 30M). Simulations run for 200K time steps unless stated otherwise. An important point of this test is that it does not use the Field solver and smoother phases. Baseline simulation: one node (Dardel), 128 MPI processes~\cite{williams2023leveraging,williams2025accelerating}.
\item \textbf{High-Density Sheath Simulation.} A double-bounded, magnetized plasma layer between two walls is considered, initially filled with electrons and $D^+$ ions. Plasma is absorbed at the walls, recycling $D^+$ ions into $D$ neutrals and forming a sheath. The 1D geometry has 3M cells, three species, and initial 200 particles per cell per charged species ($\approx$ 1.2B total). Simulations run for 100K time steps unless stated otherwise. Baseline simulation: five nodes (Dardel), 640 MPI processes, exceeding single-node memory~\cite{williams2023leveraging,williams2025accelerating}.
\end{itemize}

\noindent In this work, we simulate hybrid BIT1 on four HPC systems (Table~\ref{tab:hpc_systems_columns}) using key parameters in Table~\ref{tab:bit1_input_parameters}.

\begin{table}[!ht]
\vspace{-0.4cm} % Adjust vertical figure placement
\centering
\renewcommand{\arraystretch}{1.3}
\begin{adjustbox}{max width=\columnwidth,center}
\footnotesize
\begin{tabular}{cccc>{\columncolor{gray!20}}c}
\hline
\textbf{Characteristic} & 
\textbf{Dardel CPU/GPU} & 
\textbf{MareNostrum5 (MN5) GPP/ACC} & 
\textbf{LUMI-C / LUMI-G} & 
\textbf{Frontier (OLCF-5)} \\
\hline
HPC System  & HPE Cray EX Supercomputer & Pre-Exascale EuroHPC Supercomputer & Pre-Exascale EuroHPC Supercomputer &  HPE Cray EX Exascale supercomputer \\
Processor  & 2 × AMD EPYC Zen2 (64 cores) & 2 × Intel Sapphire Rapids 8480+ (56 cores) & 2 × AMD EPYC 7763 (64 cores) &  1 × AMD EPYC (64 cores) \\
CPU Nodes  & 1,278 & 6,480 & 2,048 &  -- \\
CPU + GPU Nodes & 62 & 1,120 & 2,978 & 9,856 \\
GPUs per Node & 4 x AMD MI250X (w/ 2 x GCDs)  & 4 × Nvidia H100 & 4 x AMD MI250X (w/ 2 x GCDs) & 4 x AMD MI250X (w/ 2 x GCDs) \\
Memory per Node & 256~GB – 2~TB & 128~GB HBM – 1~TB & 256~GB – 1~TB & 512~GB DDR4 \\
Network & Slingshot 200~GB/s & NDR200 InfiniBand & Slingshot-11 & Slingshot up to 800~Gb/s \\
Storage & 679~PB  & 248~PB Online / 402~PB Archive & 117 PB  & 679~PB \\
Location  & KTH Royal Institute of Technology & Barcelona Supercomputing Center & CSC – IT Center for Science &  Oak Ridge Leadership Computing Facility \\
\hline
\end{tabular}
\end{adjustbox}
\vspace{1mm}  % adjust vertical spacing below
\caption{Key characteristics of the HPC systems used in performance tests.}
\label{tab:hpc_systems_columns}
\vspace{-1.3cm} % Adjust vertical figure spacing
\end{table}

\begin{table}[!ht]
\vspace{-0.4cm} % Adjust vertical figure placement
\centering
\renewcommand{\arraystretch}{1.3}
\begin{adjustbox}{max width=\columnwidth,center}
\tiny
\begin{tabular}{c
                p{0.65\columnwidth} % Description
                >{\columncolor{gray!20}\centering\arraybackslash}p{0.15\columnwidth} % Minimal I/O
                >{\columncolor{gray!20}\centering\arraybackslash}p{0.15\columnwidth}} % Heavy I/O
\hline
\textbf{Parameter} & \textbf{Description} & \textbf{Minimal I/O \& Diagnostics} & \textbf{Heavy I/O \& Diagnostics} \\
\hline
\texttt{slow} & Sets the plasma profiles and distribution function diagnostics (default=0). & 0 & 0 \\
\texttt{datfile} & Enables time-averaged diagnostics of plasma and particle distributions. & 0 & 1000 \\
\texttt{dmpstep} & Defines when the simulation state is written for restart or checkpointing. & 0 & 5000 \\
\texttt{mvflag} & Specifies a diagnostic snapshot averaged over \texttt{mvflag} time steps. & 0 & 1000 \\
\texttt{mvStep} & Sets the interval between successive diagnostic outputs. & 0 (< Last\_step) & 5000 \\
\texttt{Last\_step} & Specifies the final time step at which the simulation terminates. & 2000 & 10000 \\
\texttt{origdmp} & Sets the format; 1=File (Serial) I/O \& 2=openPMD (Parallel) BP4/SST. & 1/2 & 1/2 \\
\hline
\end{tabular}
\end{adjustbox}
\vspace{1mm}
\caption{Main input parameters for simulation output and control, with key flags for Minimal and Heavy I/O \& Diagnostics.~\cite{williams2024enabling,williams2024understanding}}
\label{tab:bit1_input_parameters}
\vspace{-1.3cm} % Adjust vertical figure spacing
\end{table}

% ========================
% # Performance Results #
% ========================

\section{Performance Results}
In this work, we evaluate a portable, multi-GPU hybrid MPI\allowbreak+OpenMP BIT1 using persistent GPU memory, contiguous 1D layouts, runtime interoperability, asynchronous execution, OpenMP offloading, and integrated openPMD and ADIOS2 on Nvidia and AMD GPU architectures.

\subsection{Profiling \& Understanding the Impact of openPMD on BIT1}
% \noindent \textbf{BIT1 openPMD SST Performance \& I/O Costs Per Process.} 
We begin by utilizing \texttt{gprof}, an open-source profiling tool, to analyze execution time and identify the most frequently used functions across MPI processes. The consolidated \texttt{gprof} report provides a detailed performance analysis of the Original BIT1 with and without openPMD and the ADIOS2 backends.

\begin{figure}[!ht]
    \vspace{-0.4cm} % Adjust vertical figure placement
    \begin{center}
        \includegraphics[width=0.80\textwidth]{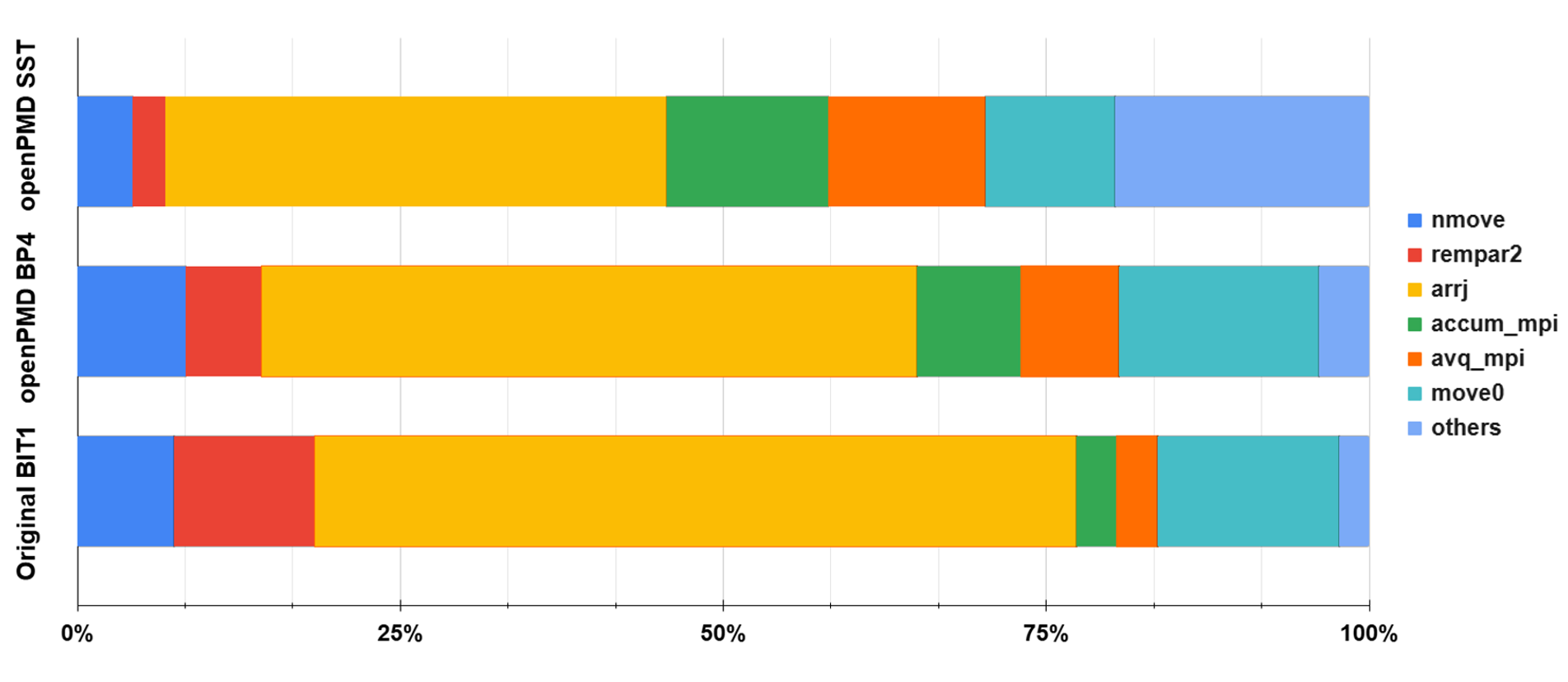}
        \caption{Ionization case function percentage breakdown (using \texttt{gprof}) on Dardel, showing where most of the execution time is spent for Original BIT1, openPMD BP4, and openPMD SST simulations~\cite{williams2023leveraging,williams2024understanding,williams2026integrating}. The \texttt{arrj} sorting function (yellow) dominates but drops from 75.5\% (Original BIT1) to 65.5\% (BP4) and 35.5\% (SST).} \label{function_breakdown}
     \end{center} 
     \vspace{-0.6cm} % Adjust vertical figure spacing
\end{figure} 

As previously reported by Williams et al.~\cite{williams2023leveraging,williams2024understanding,williams2026integrating}, Fig.~\ref{function_breakdown} shows the most time-consuming functions in the "BIT1 openPMD SST" simulation compared with the "BIT1 openPMD BP4" and "Original BIT1" simulations. In the “Original BIT1”, \texttt{arrj} (a sorting function) dominates at 75.5\%, decreasing to 65.5\% in BP4 and 35.5\% in SST, highlighting improved data handling. The particle mover \texttt{move0} drops from 18\% to 9.2\%, and the particle removal routine \texttt{rempar2} from 14\% to 2\%. The neutral particle mover \texttt{nmove} increases in BP4 but falls to 3.9\% in SST. The profiling functions \texttt{avq\_mpi} (profile computation) and \texttt{accum\_mpi} (smoothed profile computation) increase in BP4 and SST, reflecting enhanced MPI communication, while the \texttt{others} functions grows in SST, indicating redistributed computation. Overall, the "BIT1 openPMD SST" simulation spends the least amount of time in the most time-consuming functions, reducing the cost of \texttt{arrj} and \texttt{move0} compared with the "BIT1 openPMD BP4" and "Original BIT1" simulations, and improving parallel efficiency and overall performance.

\subsection{Porting \& Accelerating Hybrid BIT1 with OpenMP on 4 GPUs}
Next, we extend BIT1 by combining MPI with OpenMP in a task-based hybrid version to mitigate load imbalance and optimize resource utilization, and then port it to MN5 ACC (GPU partition) using one node with four MPI ranks and four GPUs (one GPU per rank) to analyze the impact of a contiguous 1D data layout, OpenMP target parallelism, and multi-GPU asynchronous execution.

\begin{figure}[!ht]
    \vspace{-0.4cm} % Adjust vertical figure placement
    \begin{center}
        \includegraphics[width=0.90\textwidth]{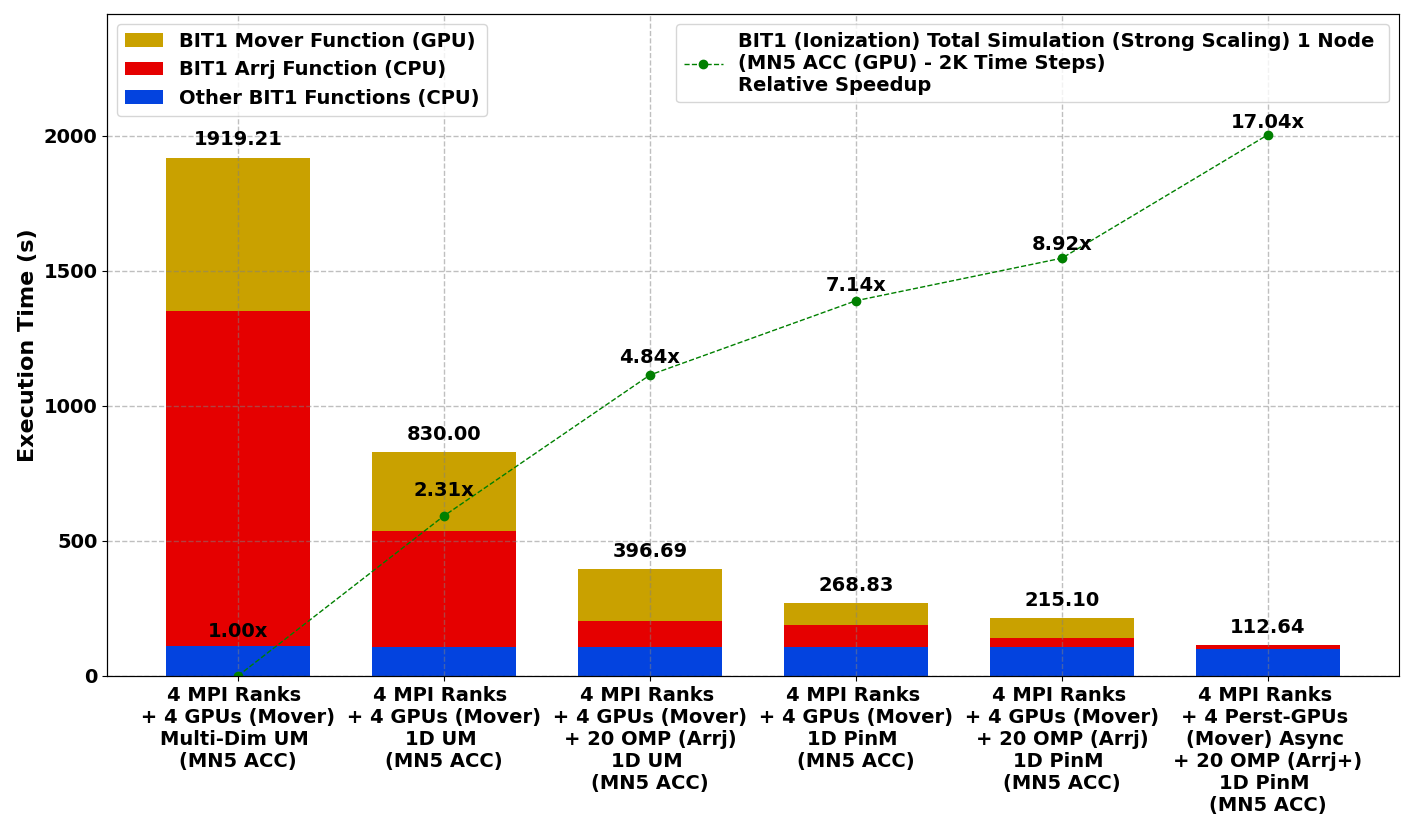}
     \end{center}
     \caption{Hybrid BIT1 (Ionization Case) Total Simulation (Development Progression) strong scaling on 1 Node (4 MPI ranks \& 4 GPUs) on MN5 ACC for 2K timesteps.} 
     \label{BIT1_Strong_Scaling_1_Node_MN5_Execution_Time_UM_and_PinM_GPU}
     \vspace{-0.6cm} % Adjust vertical figure placement
\end{figure}

As seen in Fig.~\ref{BIT1_Strong_Scaling_1_Node_MN5_Execution_Time_UM_and_PinM_GPU}, the original BIT1 version, using 4 MPI + 4 GPUs (Mover) with a 3D data layout and UM, executed with a total simulation time of $\approx$1919~s, with the \texttt{arrj} function dominating at $\approx$1242~s and the \texttt{mover} function taking $\approx$568~s. Transitioning to a 1D UM layout improves memory access, reducing the total time to $\approx$830~s and lowering the \texttt{mover} and \texttt{arrj} times to $\approx$292~s and $\approx$430~s, respectively. Introducing OpenMP thread parallelism to the \texttt{arrj} function further decreases its execution time to $\approx$94~s, bringing the total simulation time to $\approx$397~s. Using PinM for the \texttt{mover} reduces data transfer overhead, decreasing the total time to $\approx$269~s, and combining PinM with OpenMP \texttt{arrj} parallelism further reduces it to $\approx$215~s. Finally, the fully optimized version with persistent GPU memory, asynchronous \texttt{mover} execution, and OpenMP \texttt{arrj} parallelism reduces the \texttt{mover} time to near zero ($\approx$0.06~s) and \texttt{arrj} to $\approx$11.5~s, resulting in a total simulation time of around $\approx$113~s, corresponding to a 17.04× speed up over the original BIT1 (3D data layout + UM) simulation. 

\subsection{Hybrid BIT1 (Minimal I/O \& Diagnostics) up to 800 GPUs}
Moving to the high-density sheath (production-like case), we evaluate the portable, multi-GPU hybrid MPI\allowbreak+OpenMP asynchronous version of BIT1 in both strong and weak scaling tests under minimal I/O and diagnostics. This is performed using a 2K time step sheath simulation up to 100 nodes (up to 800 GPUs), with the parameters listed in Table~\ref{tab:bit1_input_parameters}, representing a near compute-only workload with a final checkpoint at the end of the simulation.

\begin{figure}[!ht]
    \vspace{-0.2cm} % Adjust vertical figure placement
    \begin{center}
        \includegraphics[width=0.92\textwidth]{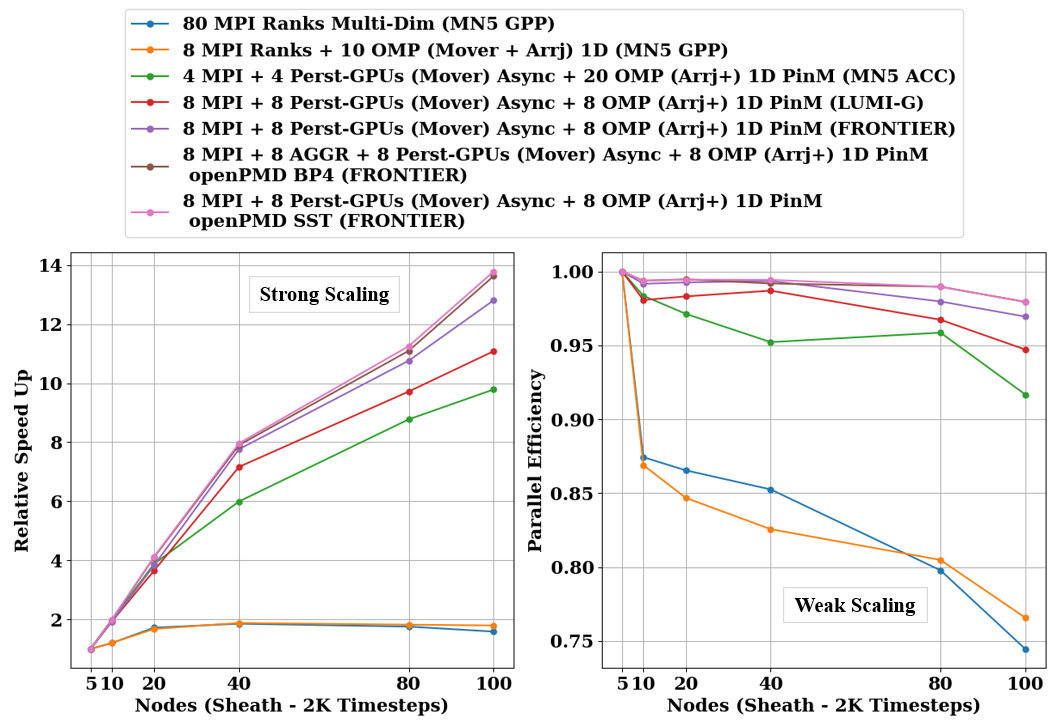}
     \end{center}
     \caption{Hybrid BIT1 (Sheath) Total Simulation (Relative) Speed Up (left) and PE (Right) - Strong and Weak Scaling up to 100 Nodes (up to 800 GPUs) on MN5 ACC, LUMI-G and Frontier for 2K timesteps.} 
     \label{BIT1_Strong_and_Weak_Scaling_100_Nodes_Frontier_Speed_Up_and_Parallel_Efficiency}
     \vspace{-0.6cm} % Adjust vertical figure spacing
\end{figure}

As seen in Fig.~\ref{BIT1_Strong_and_Weak_Scaling_100_Nodes_Frontier_Speed_Up_and_Parallel_Efficiency}, the original BIT1 version on MN5 GPP saturates under strong scaling beyond 20 nodes, with total execution times between $\approx$211~s and $\approx$247~s at 40–100 nodes, while the optimized CPU version shows only modest improvements and follows a similar trend. In contrast, the hybrid BIT1 versions exhibit substantially better scaling. On MN5 ACC, the total time decreases from $\approx$283.99~s at 5 nodes to $\approx$29.02~s at 100 nodes, corresponding to a speed up of 9.73×. On LUMI-G, execution time decreases from $\approx$175.21~s to $\approx$15.81~s (11.08× speed up). On Frontier, the fully optimized GPU version reaches $\approx$12.96~s (12.80× speed up). Enabling openPMD with ADIOS2 introduces only marginal overhead, with the BP4 and SST backends completing in $\approx$11.85~s (13.64× speed up) and SST in $\approx$11.71~s (13.81× speed up) respectively. 

For weak scaling, where the problem size increases proportionally with the number of nodes, the hybrid GPU versions maintain nearly constant execution times from 5 to 100 nodes, whereas both CPU versions show a steady increase. On Frontier, execution time remains almost flat, increasing from $\approx$165.84~s at 5 nodes to $\approx$171.06~s at 100 nodes, resulting in a parallel efficiency (PE) of 97.0\%. With openPMD BP4, execution increases from $\approx$161.45~s to $\approx$164.84~s (97.9\% PE), and with SST, from $\approx$161.36~s to $\approx$164.75~s (98.0\% PE), demonstrating that the GPU versions achieve substantial acceleration while maintaining high weak scaling efficiency under minimal I/O and diagnostics.

\subsection{Hybrid BIT1 (Heavy I/O \& Diagnostics) up to 16,000 GPUs}
We evaluate the exascale readiness of the portable, multi-GPU hybrid MPI\allowbreak+OpenMP version of BIT1 in both strong and weak scaling tests under heavy I/O and diagnostics by performing a 10K time-step sheath simulation on Frontier up to 2000 nodes (up to 16,000 GPUs) with the parameters listed in Table~\ref{tab:bit1_input_parameters}. This represents a heavier diagnostic workload with frequent time dependent output and checkpointing, which heavily stresses computation, communication, I/O, in-situ analysis and visualization, and most importantly, the file system.

\begin{figure}[!ht]
    \vspace{-0.4cm} % Adjust vertical figure placement
    \begin{center}
        \includegraphics[width=0.92\textwidth]{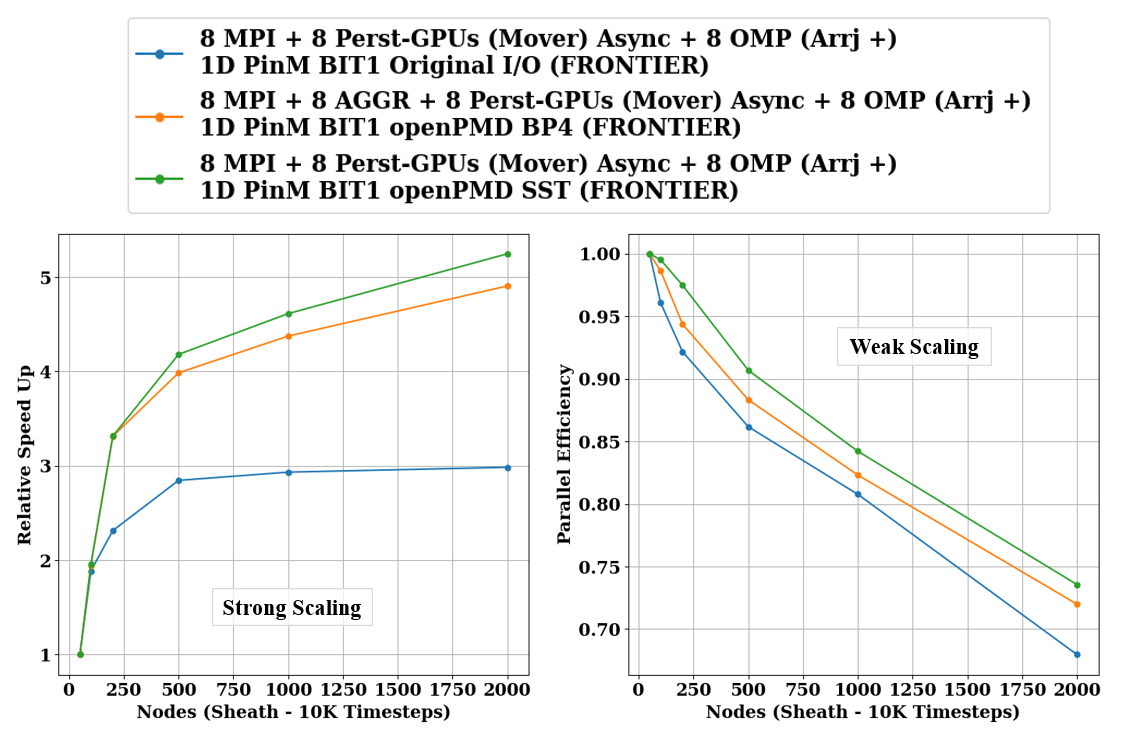}
     \end{center}
     \caption{Hybrid BIT1 (Sheath) Total Simulation (Relative) Speed Up (left) and PE (Right) - Strong and Weak Scaling up to 2000 Nodes (up to 16,000 GPUs) on Frontier for 10K timesteps.} 
     \label{BIT1_Strong_and_Weak_Scaling_2000_Nodes_Frontier_Speed_Up_and_Parallel_Efficiency}
     \vspace{-0.6cm} % Adjust vertical figure spacing
\end{figure}

As seen in Fig.~\ref{BIT1_Strong_and_Weak_Scaling_2000_Nodes_Frontier_Speed_Up_and_Parallel_Efficiency} and strong scaling tests, the original hybrid BIT1 GPU (serial I/O) version achieves a speed up of 2.99× when scaling from 50 to 2,000 nodes. In contrast, the openPMD GPU (parallel I/O) implementation with the ADIOS2 BP4 backend reaches a speed up of 4.90×, while the SST backend further improves this to 5.25× speed up. This demonstrates that both openPMD backends improves scalability under heavy I/O, with SST providing the highest scaling efficiency for large GPU runs.

For weak scaling, the corresponding PE at 2,000 nodes is 67.9\% for the original hybrid BIT1 GPU version, while the openPMD GPU version with BP4 sustains 72.0\% PE, and the openPMD GPU version with SST further improves this to 73.6\% PE. Despite the extremely heavy diagnostic and I/O workload, the openPMD GPU implementations with ADIOS2 backends (BP4 and SST) consistently maintain higher PE than the original hybrid BIT1 GPU implementation, indicating more robust weak scaling behavior and improved resilience to I/O and diagnostics pressure at scale.

% ==================
% # Discussion and Conclusion #
% ==================

\section{Discussion \& Conclusion}
This work presents a portable, multi-GPU hybrid MPI+OpenMP implementation of BIT1 for large-scale PIC MC simulations on pre-exascale and exascale supercomputing architectures. By combining persistent device-resident memory, a contiguous 1D data layout, PinM, runtime interoperability and asynchronous multi-GPU execution using OpenMP target tasks with explicit dependencies, BIT1 effectively reduces data movement and synchronization overheads while improving utilization of multiple accelerators.

On Frontier, hybrid BIT1 demonstrates good scalability under both minimal and heavy I/O and diagnostics. With minimal I/O and diagnostics up to 100 nodes (up to 800 GPUs), the fully optimized GPU version reaches $\approx$12.96 s (12.80× speed up), while openPMD BP4 and SST achieve 13.64× and 13.81× speed up, respectively, with weak scaling remaining nearly flat ($\approx$165.84–$\approx$171.06 s, 97\%–98\% PE). Under heavy I/O and diagnostics from 50 to 2000 nodes (up to 16,000 GPUs), strong scaling speed up reaches 5.07× (BP4) and 5.25× (SST), while weak scaling achieves 72.0\%–73.6\% PE, sustaining intensive I/O and frequent diagnostics, demonstrating the effectiveness of combining asynchronous multi-GPU execution with standardized data management at scale.

\begin{figure*}[!ht]
    \vspace{-0.4cm} % Adjust vertical figure placement
    \begin{center}
        \includegraphics[width=\textwidth]{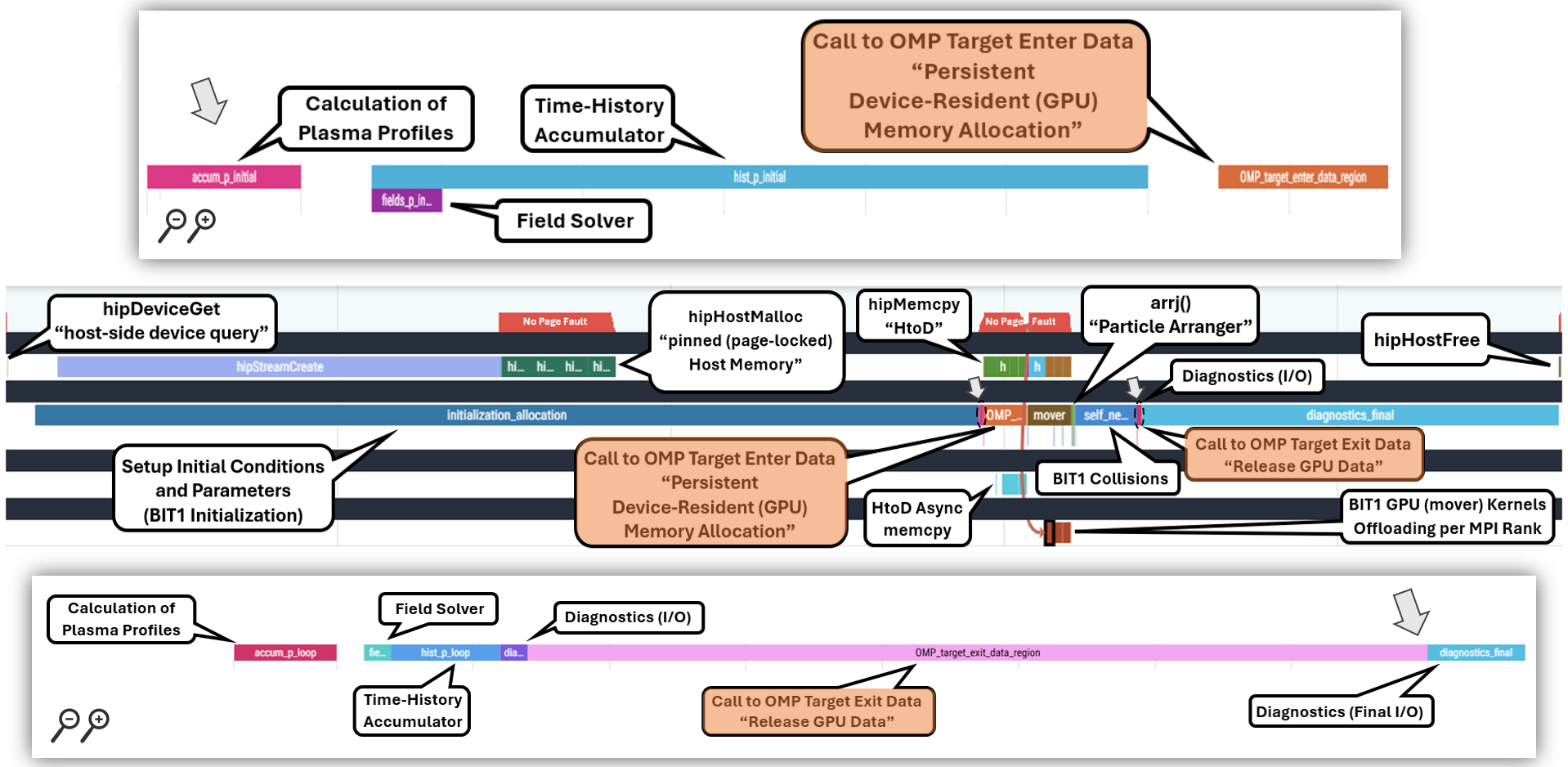}
     \end{center}
     \caption{A single time step Hybrid BIT1 AMD MI250X GPU activity trace showing HSA and HIP activity, ROCTX regions, asynchronous data copies, and mover kernel execution, obtained using \texttt{rocprof} and visualized with Perfetto on Dardel GPU, with corresponding confirmation traces on both LUMI-G and Frontier.} 
     \label{AMD_Persident_GPU_Update}
     \vspace{-0.4cm} % Adjust vertical figure spacing
\end{figure*}

Future research will extend hybrid BIT1 to Intel GPU platforms at exascale, targeting \texttt{Aurora}, and Europe’s first exascale system, \texttt{JUPITER Booster}, to assess portability across Nvidia, AMD and Intel GPUs, and to identify any remaining architecture specific limitations. Deeper performance analysis will be carried out using specialized profiling and tracing tools, including \texttt{Nvidia Nsight Systems} (Nvidia GPUs), the open-source portable profiling and tracing toolkit \texttt{Tuning and Analysis Utilities} (TAU), and AMD GPU profiling and tracing tools, such as AMD ROC-Profiler (\texttt{rocprof}). For instance, we can use \texttt{rocprof} to collect correlated traces of Heterogeneous System Architecture (HSA) and Heterogeneous-computing Interface for Portability (HIP) activity, including ROCm Tools Extension (ROCTX) regions (\texttt{roctxRangePush()} and \texttt{roctxRangePop()}), asynchronous data transfers, and GPU kernel execution, all captured in JSON format for use with Perfetto, enabling immediate visualization of key hybrid BIT1 trace activity (see Fig.~\ref{AMD_Persident_GPU_Update}). Finally, the integration of HPC workflows with AI-based methods will be investigated to further enhance real-time analysis, adaptive control, and performance optimization beyond pre-exascale and exascale supercomputing capabilities.

% \space
% ==================
% # Acknowledgment #
% ==================

% use section* for acknowledgment
%\section*{}
\vspace{2mm} 
\noindent \footnotesize{\textbf{Acknowledgments.} Funded by the European Union. This work has received funding from the European High Performance Computing Joint Undertaking (JU) and Sweden, Finland, Germany, Greece, France, Slovenia, Spain, and Czech Republic under grant agreement No 101093261 (Plasma-PEPSC). The computations/data handling were/was enabled by resources provided by the National Academic Infrastructure for Supercomputing in Sweden (NAISS), partially funded by the Swedish Research Council through grant agreement no. 2022-06725. This research used resources of the Oak Ridge Leadership Computing Facility at the Oak Ridge National Laboratory, which is supported by the Advanced Scientific Computing Research programs in the Office of Science of the U.S. Department of Energy under Contract No. DE-AC05-00OR22725

% Example citation, See \cite{lamport94}.

%% If you have bib database file and want bibtex to generate the
%% bibitems, please use
%%
%%  \bibliographystyle{elsarticle-num} 
%%  \bibliography{<your bibdatabase>}

%% else use the following coding to input the bibitems directly in the
%% TeX file.

%% Refer following link for more details about bibliography and citations.
%% https://en.wikibooks.org/wiki/LaTeX/Bibliography_Management

\bibliographystyle{splncs04}

\bibliography{bit1paper}

% \begin{thebibliography}{00}

%% For numbered reference style
%% \bibitem{label}
%% Text of bibliographic item

% \bibitem{ayguade2008design}
% Ayguadé, E., et al.: The design of openmp tasks. IEEE Transactions on Parallel
% and Distributed systems 20(3), 404–418 (2008)

% \end{thebibliography}

\end{document}